\documentclass[10pt,a4paper,twoside]{article}
\usepackage{epsfig}
\usepackage{baltlat5}
\usepackage{wrapfig}
\begin{document}
\ \
\vspace{-2.5mm}

\setcounter{page}{539}

\titlehead{Baltic Astronomy, vol.\ts 15, 539--546, 2006.}

\titleb{HIGH VELOCITY SPECTROSCOPIC BINARY ORBITS FROM\\ PHOTOELECTRIC
RADIAL VELOCITIES: BD +30 2129\,A}

\begin{authorl}
\authorb{A. Bartkevi\v cius}{1,3} and
\authorb{J. Sperauskas}{1,2}
\end{authorl}

\begin{addressl}
\addressb{1}{Institute of Theoretical Physics and Astronomy,
Vilnius University,\\ Go\v{s}tauto 12, Vilnius, LT-01108, Lithuania}

\addressb{2}{Vilnius University Observatory, \v Ciurlionio~29, Vilnius,
 LT-03100, Lithuania}

\addressb{3}{Department of Theoretical Physics, Vilnius Pedagogical
University, Student\c u~39,\\ Vilnius, LT-08106, Lithuania}

\end{addressl}

\submitb{Received 2006 November 16; accepted 2006 December 15}

\begin{summary} The spectroscopic orbit of a high proper motion visual
binary system BD +30 2129 component A is determined from 22 CORAVEL-type
radial velocity measurements.  A period of $P$ = 32.79 days and a
moderate eccentricity $e$ = 0.29 are obtained.  The visual system AB has
a projected spatial separation $\sim$\,580 AU.  The system's barycenter
velocity $V_0$ = \hbox{--35.95 km/s} and the transverse velocity $V_{\rm
t}$ = 132.2 km/s.  The Galactic spatial velocity components $U$ = +76.7
km/s, $V$ = \hbox {--110.4 km/s}, $W$ = \hbox {--26.6 km/s}, and a large
ultraviolet excess give evidence that the star belongs to thick disk
population of the Galaxy.  \end{summary}


\resthead{High velocity spectroscopic binary orbits:  BD +30 2129\,A}{A.
Bartkevi\v cius, J. Sperauskas }

\sectionb{1}{INTRODUCTION}

Some time ago we initiated a program of photoelectric measurements of
radial velocities of Population II single and binary stars
(Bartkevi\v cius et al. 1992; Sperauskas
\& Bartkevi\v cius 2002; Bartkevi\v cius \& Sperauskas 2005a,b and
references therein).  This is the second publication devoted to
spectroscopic binary orbits of high velocity stars using radial
velocities, obtained by one of us (J.S.) with a CORAVEL-type
spectrometer.

As a moderate high-proper-motion star ($\mu$ = 0.187\arcsec /yr),
BD\,+30 2129 was most probably first recorded by Luyten in his Bruce
Proper Motion catalog (Luyten 1960) as BPM 87250.  Respectively, the
star is also included in the Luyten NLTT (Luyten 1979) catalog.  As a
visual binary it was discovered by {\it Hipparcos} (ESA
1997).

\sectionb{2}{IDENTIFICATION}

Equatorial coordinates of the binary components, taken from the {\it
Hipparcos} Catalogue Double and Multiple Systems Annex (DMSA), part C
(ESA 1997), are (ICRS, epoch J1991.25):

\noindent A:  RA = 11$^{\rm h}$17$^{\rm m}$3.5272$^{\rm s}$, DEC =
+29$\degr$19\arcmin\ 26.307\arcsec,

\noindent B:  RA = 11$^{\rm h}$17$^{\rm m}$3.3515$^{\rm s}$, DEC =
+29$\degr$19\arcmin\ 29.534\arcsec.
\newpage

\begin{wrapfigure}[20]{i}[0pt]{65mm}
\vspace{-2mm}
\psfig{figure=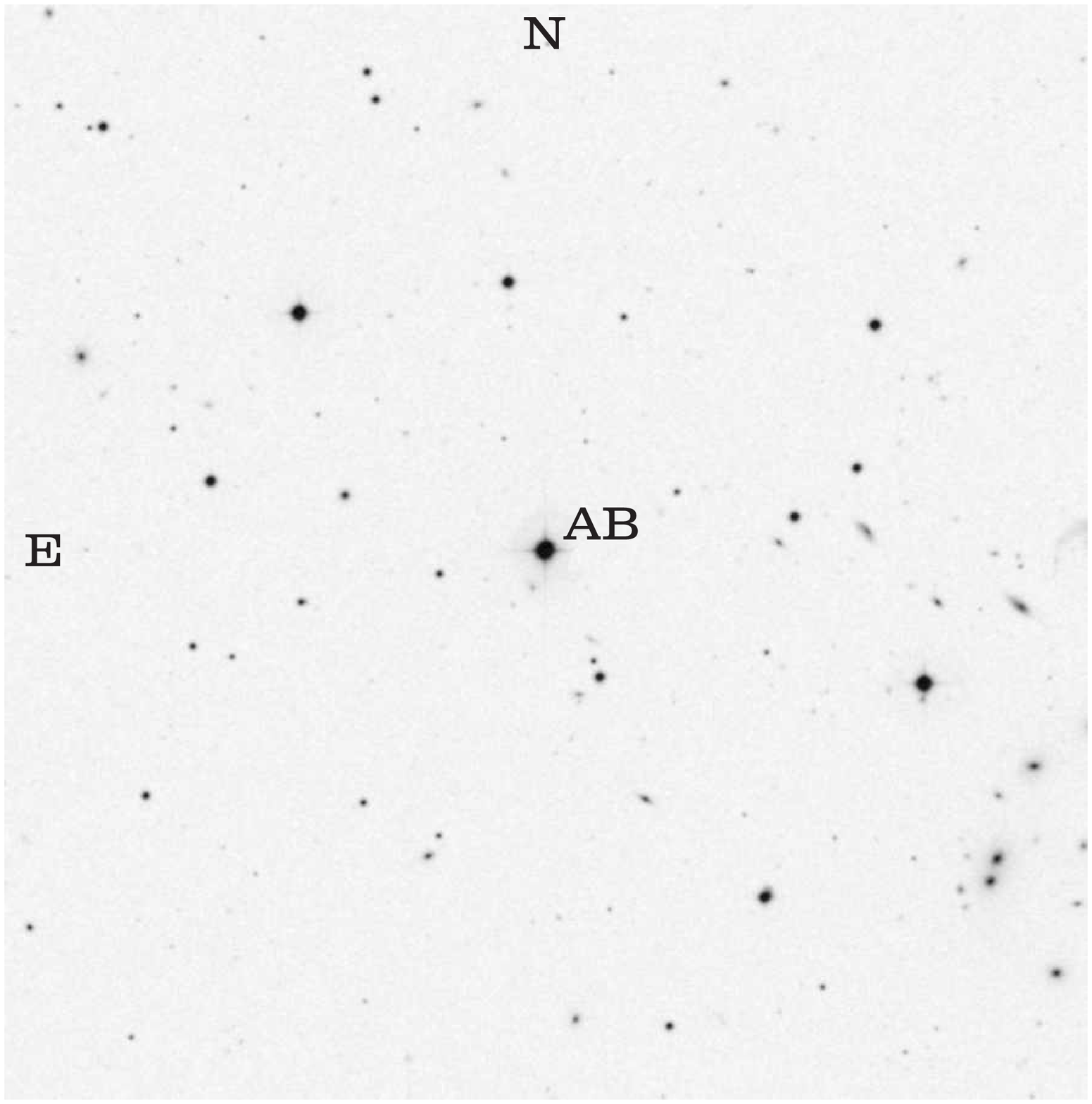,width=64mm,angle=0,clip=}
\vspace{1mm}
\captionb{1}{\hstrut Identification chart from the second blue Palomar Sky
Survey (POSS II~J). The size of the image is 13\arcmin $\times$
13\arcmin.}
\end{wrapfigure}

Due to the fact that component A is comparatively bright and the
distance between the components is only 4\arcsec, in all photographic
sky surveys and even in the 2MASS survey component B is blended by
the large and saturated image of component A.

An interesting fact is that the successive BD star in the declination
zone +30$\degr$, BD +30 2130, at the angular distance of
$\sim$\,15\arcmin,
is also a moderately high-proper-motion star (G\,147-36, $\mu$ =
0.198\arcsec / yr).  It is also a short period ($P$ = 6.57206 d)
metal-deficient spectroscopic binary (Latham et al. 2002).

The identification chart from the second blue Palomar Sky Survey
(POSS\,IIJ) is presented in Figure 1. Identifiers of the star taken from
the CDS Simbad, and retrieved using the Wizier and other sources, are
presented in Table 1. It is obvious that Simbad gives only a part of
identifiers.

\vspace{2mm}

%\begin{center}
\vbox{\footnotesize
\begin{tabular}{lll}
\multicolumn{3}{l}{\parbox{100mm}{\baselineskip=8pt
{\normbf Table 1.}{\norm\ ~Identifiers of the binary components. }}}\\
\tablerule
\noalign{\vskip2mm}

   A: BD +30 2129      &   AB: WDS J11171+2919   &   A: ASCC-2.5 683301         \\
  AB: HIP 55115        &    A: BPM 87250         &   B: ASCC-2.5 683300         \\
   A: TYC 1983-1992-1  &    A: NLTT  26848       &   A: USNO-B1.0 1193-0187375  \\
   B: TYC 1983-1992-2  &    A: FONAC 956397      &   B: USNO-B1.0 1193-0187372  \\
  AB: HDS 1608         &    A: GSC2.2 N201302235 &   A: UCAC BS 50077892        \\
   A: GSC 01983-01992  &   AB: TDSC 31153        &   B: UCAC BS 50077891        \\
  AB: CCDM J11171+2919 &                         &                              \\
\noalign{\vskip.2mm}
\tablerule
\end{tabular}
}
\vspace{-3mm}

\sectionb{3}{PHOTOMETRY }

Only one ground-based photoelectric photometry publication is known
(Figue\-ras et al. 1990).  The color indices $V$--$R$ and $V$--$I$ are
in
the Neckel \& Chini (1980) system.  Although the authors do not indicate,
both components probably are measured together, since the distance
between the components A and B is rather small.  As was mentioned
above, {\it Hipparcos} has resolved components A and B. A new reduction
in the {\it Tycho 2} catalog (H{\o}g et al. 2000) presents very
different $B_T$ and $V_T$ values for component A, in comparison to those
given in {\it Hipparcos} DMSA-C.  As {\it Tycho 2} presents more precise
and more consistent photometry with the ground-based data of Figueras et
al.  (1990), we use only the {\it Tycho 2} values for component A.
Photometry of the components estimated and reduced to the standard
systems is given in Table 2. The methods of estimation and reduction are
commented in the notes to Table 2. The components have been measured in
2MASS (Skrutskie et al. 2006), but those measurements suffered from very
strong blending and are unreliable, especially for component B.

{\it UBV} photometry indicates a considerable ultraviolet excess for the
combined light of AB, the same as it is estimated for component A, and
the main component Aa of the spectroscopic binary.  Prior to the
determination of ultraviolet excesses $\delta$($U$--$B$), color indices
were corrected for a small interstellar reddening $E_{B-V}$ $\approx$
$E_{U-B}$ $\approx$ 0.01 mag.  $\delta$($U$--$B$) is determined from the
$U$--$B$ vs.  $B$--$V$ diagram, using the mean main sequence line
(Bartkevi\v cius 1980).  For both cases, AB and A,
the normalized ultraviolet excess $\delta$($U$--$B$)$_{0.6}$,
extrapolated using the Sandage (1969) Table 1A, is 0.19 mag.  Using the
Karatas \& Schuster (2006) relation (Eq. 6) we obtain [Fe/H] = --1.1.
\vspace{-2mm}

\begin{center}
\vbox{\footnotesize
\begin{tabular}{crcrcccccc}
\multicolumn{10}{l}{\parbox{120mm}{\baselineskip=8pt
{\normbf\ \ Table 2.}{\norm\ Photometry }}}\\
\tablerule
Comp. & $B_{\rm T}$~ & $\sigma_{B_{\rm T}}$ & $V_{\rm T}$~ &$\sigma_{V_T}$&$B_T$--$V_T$ &
$\sigma_{(B_T-V_T)}$ & $H_p$ & $\sigma_{H_p}$ & Notes  \\
\tablerule
\noalign{\vskip.2mm}
 A & 10.566 &0.036 & 10.112& 0.039 & 0.454 & 0.053  & 10.151& 0.015 & 1\\
 B &        &      &       &       &       &        & 13.133& 0.234 & 1\\
 A & 10.625 &0.040 & 9.992 & 0.033 & 0.633 & 0.052  &       &       & 2\\
\tablerule
\end{tabular}
}
\end{center}
\vspace{-9mm}

\begin{center}
\vbox{\footnotesize
\tabcolsep = 4pt
\begin{tabular}{cccccccc}
\multicolumn{8}{c}{\parbox{120mm}{\baselineskip=8pt {}}}\\
\tablerule
  Comp. & $B$ & $\sigma_B$ & $V$ & $\sigma_V$ & $B$--$V$  &$\sigma_{(B-V)}$ &  Notes  \\
\tablerule
\noalign{\vskip.2mm}
A  & 10.474 & 0.033 & 9.927 &0.032 &  0.547 & 0.046 &  3 \\
B  &        &       &\llap{1}3.126 &0.232 &    &  & 3 \\
\tablerule
\end{tabular}
}
\end{center}
\vspace{-9mm}

\begin{center}
\vbox{\footnotesize
\tabcolsep = 4pt
\begin{tabular}{cccccccc}
\multicolumn{8}{c}{\parbox{120mm}{\baselineskip=8pt {}}}\\
\tablerule
 Comp. & $V$ & $\sigma_V$ & $B$--$V$  &  $\sigma_{(B-V)}$ & $U$--$B$ &$\sigma_{(U-B)}$ & Notes\\
\tablerule
\noalign{\vskip.2mm}
AB  & 9.86\rlap{0} &0.015 & 0.57\rlap{3} & 0.034 & --0.08\rlap{7} &0.019 & 4 \\
 A  & 9.93  &      & 0.56  &       & --0.10  &       & 5  \\
Aa  & \llap{1}0.26  &      & 0.52  &       & --0.18  &       & 6a \\
Aa  & 9.94  &      & 0.53  &       & --0.12  &       & 6b \\
 B  & \llap{1}3.04  &      &       &       &        &       & 7 \\
\tablerule
\end{tabular}
}
\end{center}
\vspace{-9mm}

\begin{center}
\vbox{\footnotesize
\tabcolsep = 4pt
\begin{tabular}{cccccccc}
\multicolumn{8}{c}{\parbox{120mm}{\baselineskip=8pt {}}}\\
\tablerule
 Comp. & $(V-R)_{\rm NC}$ & $\sigma_{(V-R)_{\rm NC}}$ & $(V-I)_{\rm NC}$ &  $\sigma_{(V-I)_{\rm NC}}$
& $(R-I)_{\rm NC}$ & $\sigma_{(R-I)_{\rm NC}}$  & Notes  \\
\tablerule
\noalign{\vskip.2mm}
AB & 0.50\rlap{7}  & 0.031 & 0.89\rlap{9}  & 0.014 &  0.39\rlap{2} &0.034 & 4 \\
A & 0.47   &      & 0.83   &      &   0.36    &      & 5 \\
\tablerule
\end{tabular}
}
\end{center}
\vspace{-9mm}

\begin{center}
\vbox{\footnotesize
\tabcolsep = 4pt
\begin{tabular}{ccccc}
\multicolumn{5}{c}{\parbox{120mm}{\baselineskip=8pt {}}}\\
\tablerule
 Comp. & ($V$--$R$)$_{\rm C}$ & ($V$--$I$)$_{\rm C}$ &  ($R$--$I$)$_{\rm C}$ &  Notes  \\
\tablerule
\noalign{\vskip.2mm}
AB &0.35  &0.70  &0.35  & 8   \\
 A &0.32  &0.64  &0.32  & 5   \\
Aa &0.30  &0.61  &0.31  & 6a  \\
Aa &0.30  &0.60  &0.30  & 6b  \\
\tablerule
\end{tabular}
}
\end{center}
\vspace{-2mm}

\noindent {\bf Notes to Table 2.}
\vskip1mm
{\footnotesize
\noindent 1. Hipparcos DMSA-C;

\noindent 2. Tycho 2;

\noindent 3. ASCC-2.5 (Kharchenko 2001);

\noindent 4. Figueras et al. (1990);

\noindent 5. $V$ and $B$---$V$ of component A are obtained by averaging:
(a)
component A photometry obtained from ground-based combined AB photometry
(Figueras et al. 1990), assuming $\Delta V{(\rm AB)}$ = 3.1 mag and
using a F7\,V spectral type for component A and a K5\,V spectral
type for component B; energy distributions were taken from
Sviderskien\.{e} (1988) and intrinsic color indices from Strai\v zys
(1992), (b) the values of $V$ and $B$--$V$ of component A published in
ASCC-2.5, (c) the values of $V$ and $B$--$V$ reduced from {\it Tycho 2}
$B_T$ and $V_T$ (using the {\it Hipparcos} equations [1.3.33] and
[1.3.26]), (d) the same as in (c), but using reductions from Table 2 of
Bessell (2000).  The color indices $U$--$B$, ($V$--$R$)$_{\rm NC}$ and
($V$--$I$)$_{\rm NC}$ of component A are obtained using only by the
above condition (a).  Here NC means color indices from Neckel \& Chini
(1980).

\noindent 6. Very approximate magnitudes and color indices of
spectroscopic binary
primary component Aa are obtained from component A photometry, assuming
(a) $\Delta V$\,(AaAb) = 1.1 mag, F6\,V spectral type for component Aa
and G5\,V type for component Ab, or (b) $\Delta V$\,(AaAb) = 4.1 mag,
F6\,V spectral type for component Aa and K7\,V type for component Ab.

\noindent 7. $V$ magnitude for component B is obtained by averaging the
value from ASCC-2.5 and the value reduced from $H_p$ magnitude, using
the {\it Hipparcos} and Bessell (2000) relations.

\noindent 8. Reduction of Neckel \& Chini (1980) color indices
($V$--$R$)$_{\rm
NC}$ and ($V$--$I$)$_{\rm NC}$ to the Cousins (1980) system was done
by the Taylor et al.  (1989) relations.  }

\sectionb{4}{DISTANCE, ABSOLUTE MAGNITUDES AND KINEMATICS }

For the star {\it Hipparcos} recorded a parallax of moderate precision
(37\%), $\pi$ = 6.88$\pm$2.56 mas.  This corresponds to a distance to
the system of $r$ = 145$\pm$54 pc.  Interstellar reddening was obtained
from the Schlegel et al.  (1998) maps, using the NED database extinction
calculator.  For BD +30 2129 at $\ell$ = 201.4$\degr$ and $b$ =
+69.1$\degr$, we got the total line-of-sight reddening $E_{B-V}$ = 0.018
mag.  This reddening was reduced by the exponential law taking into
account the Galactic latitude and the distance (as described by
Anthony-Twarog \& Twarog 1994, p. 1583), and the values $E_{B-V}$ =
0.012 and $A_V$ = 0.04 were obtained for the star.  Adopting the {\it
Hipparcos} parallax, $A_V$ = 0.04, $V_0$(Aa) = 10.0 mag and $V_0$(B) =
13.0 mag, we obtain absolute magnitudes $M_V$ = 4.2$\pm$0.8 and
7.2$\pm$0.8 for A and B components.  These absolute magnitudes
correspond to spectral types F8\,V and K5\,V for solar chemical
composition stars (according to the calibration from Strai\v zys 1992,
App. 2).  The color index $B$--$V$ for component A from Table 2, with a
small reddening correction, is also consistent with the F8\,V spectral
class.  Using the results of photometry and the mean weighted proper
motion value $\mu$(A+B) = 0.19188$\pm$0.00054\arcsec\ per year (taken
from ASCC-2.5), for components A and B we obtain the following
approximate reduced proper motion values $H_V$:  11.4 and 14.4.  These
$H_V$ and the estimated $(B-V)_0$ values place both components in the
subdwarf region, close to the subdwarf and the main sequence border.

The {\it Hipparcos} parallax, proper motion components
from ASCC-2.5 and our value of spectroscopic binary barycenter radial
velocity were used to calculate kinematical parameters of the system.
The procedure of computation is the same as in Bartkevi\v cius \& Gudas
(2001, 2002).  The velocity component $U$ is directed to the Galactic
center, $V$ -- to the direction of Galactic rotation and $W$ -- to the
North Galactic Pole.  The velocity components were corrected due to the
Solar motion with respect to the Local Standard of Rest $U$ =
10.0$\pm$0.4 km/s, $V$ = 5.2$\pm$0.6 km/s and $W$ = 7.2$\pm$0.4 km/s
(Binney \& Merrifield 1998).  The calculated kinematical parameters and
the data used for their calculations are given in Table 3. Evidently,
the binary system belongs to the thick disk population.
\vspace{-2mm}

\vbox{
\begin{center}
\vbox{\footnotesize
\begin{tabular}{ccccccccc}
\multicolumn{9}{l}{\parbox{120mm}{\baselineskip=8pt
 {\normbf\ \ Table 3.}{\norm\ Kinematical parameters and the data used
for their calculation.}}}\\
\tablerule
\noalign{\vskip.2mm}
$\ell$ & $b$ & $\pi$ & $\sigma_\pi$ & $\sigma_\pi/\pi$ & $\mu$\,(RA) &
$\sigma_\mu$\,(RA) & $\mu$\,(DEC) & $\sigma_\mu$\,(DEC)  \\
deg & deg & \arcsec & \arcsec &    & \arcsec/yr &  \arcsec/yr &
\arcsec/yr & \arcsec/yr  \\
201.35 & 69.13 & 0.00688 & 0.00256 & 0.372 & 0.02763 & 0.00035 &
--0.18988 & 0.00059  \\
\tablerule
\noalign{\vskip.2mm}
$\mu$ & $\sigma_\mu$ & $\sigma_\mu/\mu$ & $V_{\rm r}$ & $\sigma V_{\rm
r}$ & $V_{\rm t}$ & $\sigma {V_{\rm t}}$ &
$U$ & $\sigma_U$  \\
\arcsec/yr & \arcsec/yr &  & km/s & km/s & km/s  & km/s & km/s & km/s \\
0.19188 & 0.00059 & 0.003 & --35.95 & 0.13 & 132.2 & 49.2 & 76.7 & 15.0
\\
\tablerule
\end{tabular}
}
\end{center}
\vspace{-8mm}

\begin{center}
\vbox{\footnotesize
\begin{tabular}{cccccc}
\multicolumn{6}{l}{\parbox{80mm}{\baselineskip=8pt}}\\
\tablerule
\noalign{\vskip.2mm}
$V$ & $\sigma_V$ & $W$ & $\sigma_W$ & $V_{\rm tot}$ & $\sigma_{V_{\rm
tot}}$ \\
km/s & km/s & km/s & km/s & km/s & km/s  \\
--110.4 & 13.7 & --26.6 & 0.7 & 137.0 & 47.5  \\
\tablerule
\end{tabular}
}
\end{center}
}

\sectionb{5}{RADIAL VELOCITY MEASUREMENTS}

Radial velocities were measured by one of the authors (J.S.)
with a CORAVEL-type spectrometer constructed at the Vilnius University
Observatory.  A description of the spectrometer, the measurements and
data reduction procedures are presented in Upgren, Sperauskas \& Boyle
(2002). 22 individual radial velocities for BD +30 2129\,A were obtained
during four observing runs.  During the first run (2000 March 15, JD
2451618 -- 2000 April 24, JD 2451658) the 2.3~m, 1.53~m and 1.5~m
telescopes of the Steward Observatory were used.  During the second run
(2002 April 4, JD 2452369 -- 2002 May 9, JD 2452404), the third run (two
measurements at 2003 May 26 and 28, JD 2452786 and JD 2452788) and the
fourth run (2006 March 6, JD 2453800 -- 2006 May 10, JD 2453866) the
1.65~m and 0.63~m telescopes of the Mol\.e\-tai Observatory were used.
These measurements were spread over the period of 2248 days.  Standard
single-measurement errors range from 0.7 to 1.1 km/s, with a mean value
of 0.9 km/s.  Individual radial velocity measurements are listed in
Table 4 together with the Heliocentric Julian Days and phases,
calculated from the orbital elements, measurement errors and residuals.

\begin{center}
\vbox{\footnotesize
\tabcolsep = 4pt
\begin{tabular}{ccccr|ccccr}
\multicolumn{10}{c}{\parbox{80mm}{\baselineskip=8pt
{\normbf\ \ Table 4.}{\norm\ Radial velocity measurements.}}}\\
\tablerule
HJD 24+ & Phase & $V_r$ & $\sigma_{V_r}$ & $O$--$C$~ & HJD 24+ & Phase
& $V_r$ & $\sigma_{V_r}$ & $O$--$C$~ \\
   &        &  km/s &  km/s & km/s &     &       &  km/s & km/s & km/s \\
\tablerule
\noalign{\vskip1mm}

51618.867  &  0.561  &    --46.0 &  1.0  &   --1.9  &  ~~52385.310 &   0.935 &     --49.4 &  0.9  &   --0.1  \\
51633.754  &  0.015  &    --27.5 &  0.9  &   --0.8  &  ~~52386.325 &   0.966 &     --41.1 &  0.9  &     0.2  \\
51635.680  &  0.074  &    --12.2 &  0.8  &     1.3  &  ~~52398.415 &   0.335 &     --22.0 &  0.8  &     0.9  \\
51657.695  &  0.745  &    --57.5 &  0.8  &     0.8  &  ~~52399.408 &   0.365 &     --24.3 &  0.8  &     1.5  \\
51658.629  &  0.774  &    --59.4 &  0.7  &     0.3  &  ~~52403.348 &   0.485 &     --36.8 &  0.9  &     0.4  \\
52369.440  &  0.451  &    --34.6 &  0.8  &   --0.5  &  ~~52404.441 &   0.519 &     --39.9 &  0.9  &     0.4  \\
52375.395  &  0.633  &    --50.6 &  0.8  &   --0.3  &  ~~52786.368 &   0.166 &     --10.6 &  1.0  &   --1.0  \\
52376.431  &  0.664  &    --52.9 &  0.8  &   --0.1  &  ~~52788.383 &   0.228 &     --14.2 &  1.0  &   --1.0  \\
52377.371  &  0.693  &    --55.2 &  0.8  &   --0.2  &  ~~53800.593 &   0.097 &     --10.3 &  0.8  &     0.4  \\
52382.361  &  0.845  &    --60.4 &  0.9  &   --0.1  &  ~~53865.381 &   0.073 &     --13.7 &  1.1  &   --0.1  \\
52383.385  &  0.876  &    --58.5 &  0.9  &   --0.1  &  ~~53866.377 &   0.103 &     --10.4 &  0.8  &   --0.1  \\

\tablerule
\end{tabular}
}
\end{center}

\begin{figure}
\centerline{\psfig{figure=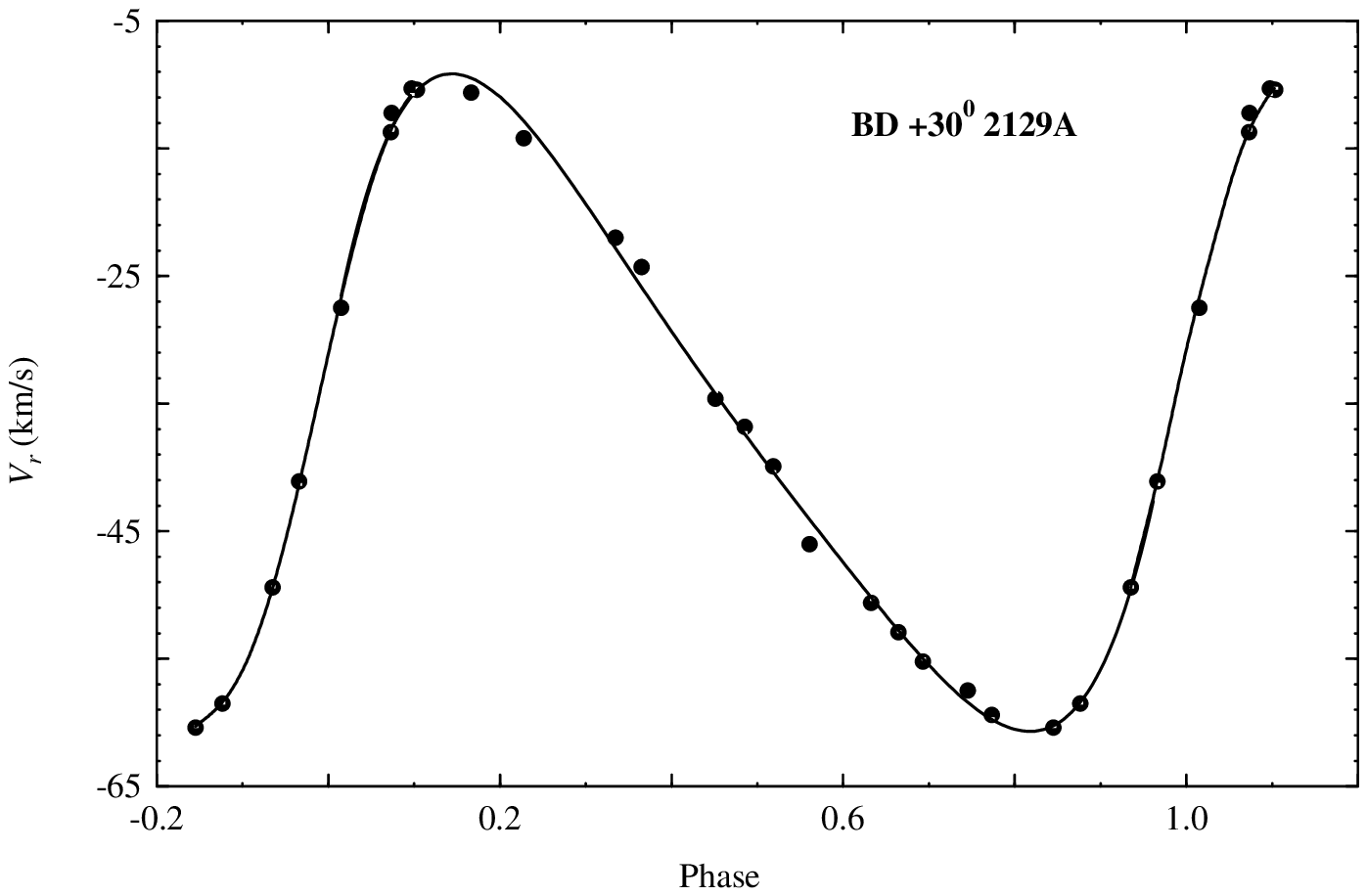,width=100mm,angle=0,clip=}}
\captionc{2}{ Radial velocity curve.}
\end{figure}

\sectionb{6}{ORBIT SOLUTION}

The obtained radial velocity curve is plotted in Figure 2. The
calculated orbital elements are given in Table 5. The system has a
moderate center-of-mass radial velocity and orbit eccentricity.

\begin{center}
\vbox{\footnotesize
\begin{tabular}{ll}
\multicolumn{2}{c}{\parbox{80mm}{\baselineskip=8pt
{\normbf\ \ Table 5.}{\norm\ Orbital elements of BD +30 2129A.}}}\\
\tablerule
Parameter & Value  \\
\tablerule
\noalign{\vskip1mm}
        Orbital period  &          $P$  =  32.790$\pm$0.002 days \\
        Center-of-mass velocity & $V_0$ =  --35.95$\pm$0.13 km/s     \\
        Half-amplitude          &  $K$ =   25.8$\pm$0.2 km/s       \\
        Eccentricity           &   $e$ = 0.289$\pm$0.007            \\
        Longitude of periastron & $\omega$ = (278.5$\pm$1.7)$\degr$ \\
Date of conjunction & $T_{\rm conj}$ = 2453863.0$\pm$0.2 HJD   \\
Projected semimajor axis &  $a \sin i$ = (11.15$\pm$0.08)\,10$^6$ km     \\
Function of the mass &  $f(m)$ = 0.0513$\pm$0.0001 $M_{\odot}$    \\
Mean square error of one observation & $\sigma_{(O-C)} = \pm$ 0.60 km/s \\
\tablerule

\end{tabular}
}
\end{center}

\sectionb{7}{PARAMETERS  OF THE VISUAL BINARY SYSTEM}

As was mentioned in Introduction, BD +30 2129, as a visual binary
system, was discovered by {\it Hipparcos} and included in the {\it
Hipparcos} Catalogue Double and Multiple Systems Annex (DMSA), part C
(systems resolved into distinct components) (ESA 1997).  This catalog
gives the angular distance $d$ = 3.962$\arcsec$ between the components
and the position angle {\it PA} = 325.0$\degr$ at epoch 1991.25.
However this is marked as an ambiguous double-star solution.  An
alternative solution for AB is also given:  {\it PA} = 16$\degr$ and $d$
= 0.29\arcsec.  The new reductions in the {\it Tycho 2} catalog, the
measurements in the 2MASS infrared survey and the visual observations
during radial velocity measurements, all definitely confirm the results
of principal solution.  The mean values using the {\it Tycho}, {\it
Tycho 2} and 2MASS solutions are:  $PA$ = 324.7$\degr$ and $d$ =
4.00\arcsec.  The minimum spatial distance between components A and B,
adopting the apparent separation $d$ = 4.00\arcsec\ and the {\it
Hipparcos} parallax $\pi$ = 0.00688\arcsec, is 581 AU.  The period of
the AB system, estimated by Kepler's third law, assuming circular
face-on orbit, apparent separation and parallax is 14\,000 or 8400
years.  The first value of the period is for the accepted mass sum of
the triple system 1 $M_{\odot}$ (this corresponds to a metal-deficient
system) and the second value is for 2.8 $M_{\odot}$ (this corresponds to
a solar chemical composition system). The mean radial velocity of
component B, $V_r$ = --34.8$\pm$1.0 km/s, based on two measurements,
confirms a binary nature of the visual system.

\sectionb{8}{CONCLUSIONS}

On the ground of 22 CORAVEL-type radial velocity measurements, a
spectroscopic orbit of moderate eccentricity ($e$ = 0.29) with a period
of $P$ = 32.79 days is obtained for component A of a high velocity
($V_{\rm tot}$= 137.0 km/s) visual binary system BD +30 2129.  The
projected spatial separation of components of the visual binary AB is
found to be $\sim$\,580 AU.  Likely, the system is metal-deficient
([Fe/H]\,$\approx$\,--1).  The Galactic spatial velocity components $U$
= +76.7 km/s, $V$ = --110.4 km/s , $W$ = --26.6 km/s, and a large
ultraviolet excess give evidence that the system belongs to thick disk
of the Galaxy.  The estimated period of the visual system is about
10\,000 years.

\thanks{ We are thankful to V.-D.  Bartkevi\v cien\. e for
preparation of the manuscript for publication.  In this investigation
information from the Strasbourg Stellar Data Center (CDS), NASA
Bibliographic Data Center (ADS), the NED database, Astrophysics preprint
archive and the Washington Visual Double Stars Catalog (WDS) was used.
Radial velocity observations were obtained with the 0.63 m and 1.65 m
telescopes of the Mol\.e\-tai Observatory, Lithuania, and the 1.5 m,
1.53 m and 2.3 m telescopes of the Steward Observatory, University of
Arizona. J. Sperauskas thanks the Steward Observatory for
providing observing time on  Kitt Peak,
Mt. Lemmon and Mt. Bigelow.}

\References
\enlargethispage{4mm}

\refb Anthony-Twarog B. J., Twarog B. A. 1994, AJ, 107, 1577

\refb Bartkevi\v cius A. 1980.  {\it The Catalogue of Metal-Deficient
F-M Stars.  Part 1. The Stars Classified Spectroscopically.} Bull.
Vilnius Obs., No. 51, 3

\refb Bartkevi\v cius A., Gudas A. 2001, Baltic Astronomy, 10, 481

\refb Bartkevi\v cius A., Gudas A. 2002, Baltic Astronomy, 11, 153

\refb Bartkevi\v cius A., Sperauskas J. 2005a, Baltic Astronomy, 14, 511

\refb Bartkevi\v cius A., Sperauskas J. 2005b, Baltic Astronomy, 14, 527

\refb Bartkevi\v cius A., Sperauskas J., Rastorguev A. S., Tokovinin
A. A. 1992, Baltic Astronomy, 1, 47

\refb Bessell M. S. 2000. PASP, 112, 961

\refb Binney J., Merrifield M. 1998  {\it Galactic Astronomy},
Princeton University Press, Princeton

\refb Cousins A.\,W.\,J. 1980, Circ.  South African Astron.  Obs, 1,
No.5, 234

\refb Figueras F., Jordi C., Rossello G., Torra J. 1990, A\&AS, 82, 57

\refb ESA, 1997, {\it The Hipparcos and Tycho Catalogues, Double and
Multiple Systems Annex (DMSA, Section C)}, ESA-SP-1200, Noordwijk

\refb H{\o}g E., Fabricius C., Makarov V. V., Urban S., Corbin T.,
Wycoff G., Bastian U., Schwekendiek P., Wicenec A. 2000.  {\it Tycho-2
Catalogue of the 2.5 Million Brightest Stars},  A\&A, 355, L27;  CDS
Catalog No.  I/259

\refb Karatas Y., Schuster W. J. 2006, MNRAS, 371, 1793

\refb Kharchenko N. V. 2001, {\it All-sky Compiled Catalogue of 2.5
Million Stars} (ASCC-2.5), Kinematics and Physics of Celestial Bodies,
Kiev, 17, 409; CDS Catalog No.  I/280A

\refb Latham D. W., Stefanik R. P., Torres G., Davis R. J., Mazeh T.,
Carney B. W., Laird J. B., Morse J. A. 2002, AJ, 124, 1144

\refb Luyten W. J. 1960.  {\it Bruce Proper Motion Survey (BPM).
Section G. Proper Motion for 14218 Stars North of Declination 0$\degr$}.
Obs.  Univ.  Minnesota

\refb Luyten W. J. 1979.  {\it NLTT Catalogue.  Vol. I. +90 to +30},
University of Minnesota, Minneapolis; CDS Catalog No. I/98A

\refb Neckel Th., Chini R. 1980, A\&AS, 39, 411

\refb Sandage A. R. 1969, ApJ, 158, 1115

\refb Schlegel D. J., Finkbeiner D. P., Davis M. 1998, ApJ, 500, 525

\refb Skrutskie M. F., Cutri R. M., Stiening R., Weinberg M. D. et al,
2006, AJ, 131, 1163.

\refb Sperauskas J., Bartkevi\v cius A. 2002, AN, 323, 139

\refb Strai\v zys V. 1992, {\it Multicolor Stellar Photometry}, Pachart
Publihing House, Tucson, Arizona

\refb Sviderskien\. e Z. 1988, Bull. Vilnius Obs., No. 80, 3

\refb Taylor B. J., Joner M. D., Johnson S. B. 1989, AJ, 97, 1798

\refb Upgren A. R., Sperauskas J., Boyle R. P. 2002, Baltic Astronomy,
11, 91

\end{document}